\documentclass[reprint,superscriptaddress,10pt,a4paper,preprintnumbers,nofootinbib]{revtex4-1}
\usepackage{amsmath,amssymb,latexsym}
\usepackage{bm}
\usepackage{graphicx}
\usepackage{hepunits}
\usepackage[svgnames]{xcolor}

\newcommand{\ff}{f\!\!f}

\newcommand{\nn}{\nonumber}

\begin{document}

\title{Invariant-mass distribution of top-quark pairs and top-quark mass determination}

\author{Wan-Li Ju}
\affiliation{School of Physics and State Key Laboratory of Nuclear Physics and Technology, Peking University, Beijing 100871, China}
\affiliation{Institute for Particle Physics Phenomenology, Durham University, Durham DH1 3LE, UK}
\author{Guoxing Wang}
\affiliation{School of Physics and State Key Laboratory of Nuclear Physics and Technology, Peking University, Beijing 100871, China}
\author{Xing Wang}
\affiliation{PRISMA+ Cluster of Excellence \& Mainz Institute for Theoretical Physics, Johannes Gutenberg University, 55099 Mainz, Germany}
\author{Xiaofeng Xu}
\affiliation{Institut f\"ur Theoretische Physik, Universit\"at Bern, Sidlerstrasse 5, CH-3012 Bern, Switzerland}
\author{Yongqi Xu}
\affiliation{School of Physics and State Key Laboratory of Nuclear Physics and Technology, Peking University, Beijing 100871, China}
\author{Li Lin Yang}
\email{yanglilin@zju.edu.cn}
\affiliation{Zhejiang Institute of Modern Physics, Department of Physics, Zhejiang University, Hangzhou 310027, China}

\begin{abstract}

We investigate the invariant-mass distribution of top-quark pairs near the $2m_t$ threshold, which has strong impact on the determination of the top-quark mass $m_t$. We show that higher-order non-relativistic corrections lead to large contributions which are not included in the state-of-the-art theoretical predictions. We derive a factorization formula to resum such corrections to all orders in the strong-coupling, and calculate necessary ingredients to perform the resummation at next-to-leading power. We combine the resummation with fixed-order results and present phenomenologically relevant numeric results. We find that the resummation effect significantly enhances the differential cross section in the threshold region, and makes the theoretical prediction more compatible with experimental data. We estimate that using our prediction in the determination of $m_t$ will lead to a value closer to the result of direct measurement.

\end{abstract}


\maketitle

\section{Introduction}

Top-quark pair production $pp \to t(p_t) + \bar{t}(p_{\bar{t}}) + X$ is one of the most important scattering processes at the Large Hadron Collider (LHC). The experimental measurements of its differential cross sections have achieved remarkably high precisions \cite{Aaboud:2018eqg, Sirunyan:2018ucr, Sirunyan:2018wem, Sirunyan:2019lnl} in the Run 2 of the LHC with a center-of-mass energy $\sqrt{s} = \unit{13}{\TeV}$. In the meantime, the theoretical modeling of the kinematic distributions in quantum chromodynamics (QCD) has also been greatly improved with the emergence of the next-to-next-to-leading order (NNLO) predictions \cite{Czakon:2015owf, Czakon:2016dgf, Catani:2019hip} supplemented with the resummation of large logarithms to the next-to-next-to-leading logarithmic accuracy (NNLL$'$) \cite{Pecjak:2016nee, Czakon:2018nun, Pecjak:2018lif}. The complete next-to-leading order (NLO) corrections incorporating electroweak (EW) effects \cite{Bernreuther:2010ny, Pagani:2016caq} have also been combined with the higher order QCD corrections, resulting in the state-of-the-art standard model (SM) predictions for this process \cite{Czakon:2017wor, Czakon:2019txp}.

\begin{figure}[t!]
\centering
\includegraphics[width=0.9\linewidth]{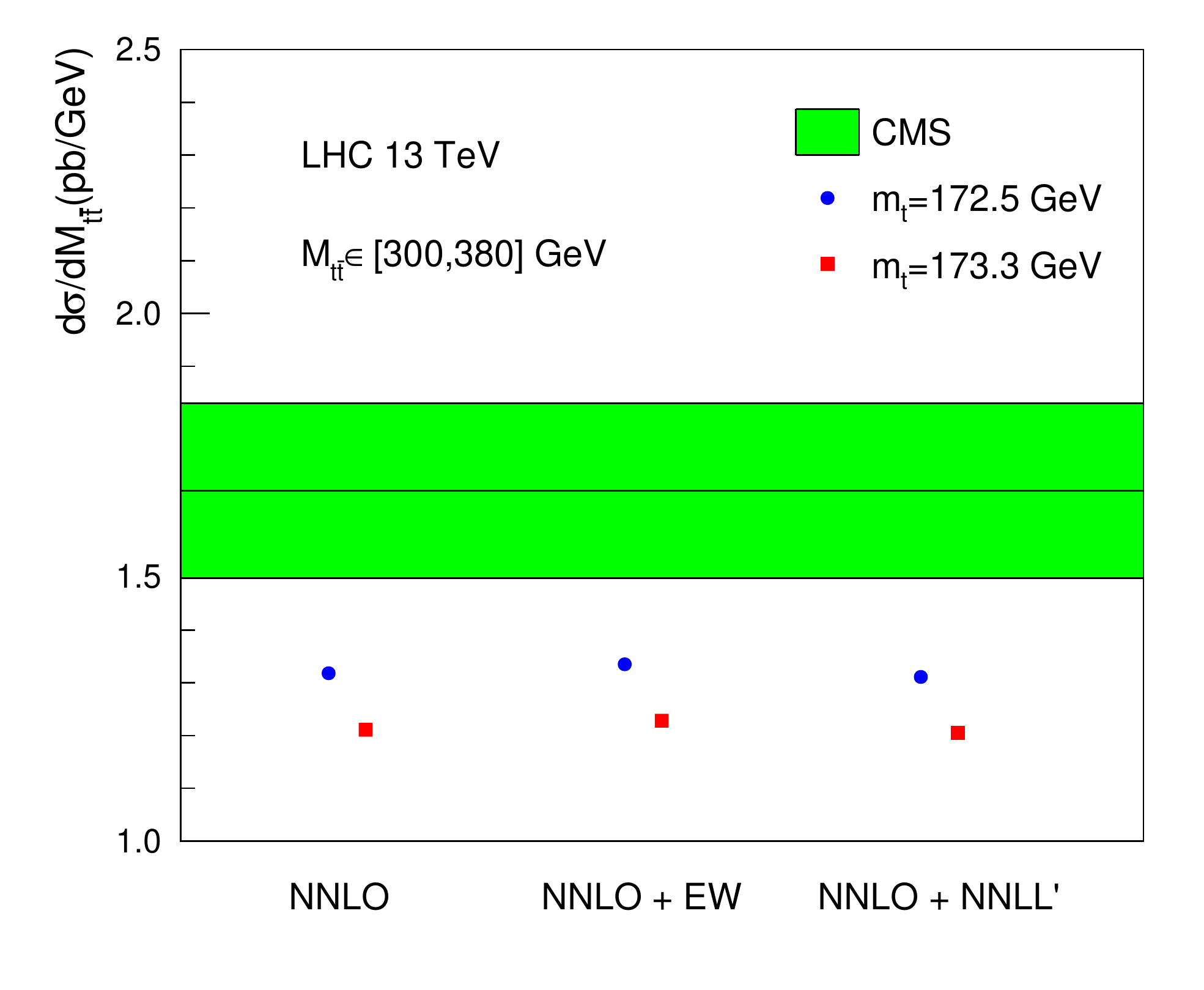}
\vspace{-5ex}
\caption{\label{fig:old_results}The averaged $t\bar{t}$ invariant mass distribution in the \unit{[300,380]}{\GeV} range. The CMS result \cite{Sirunyan:2018ucr} is shown as the green band, while the central-values of various theoretical predictions are shown in comparison.}
\end{figure}

In spite of these developments, some small discrepancies between the high precision theoretical and experimental results are persistent. A notable one is the $t\bar{t}$ invariant-mass distribution near the $2m_t$ threshold measured by the CMS collaboration at the \unit{13}{\TeV} LHC \cite{Sirunyan:2018wem, Sirunyan:2018ucr}, where $m_t$ is the top-quark mass, and the invariant mass $M_{t\bar{t}}$ is defined as $M_{t\bar{t}}^2 \equiv (p_t + p_{\bar{t}})^2$.
In Figure~\ref{fig:old_results}, we depict the averaged $M_{t\bar{t}}$ distribution in the \unit{[300,380]}{\GeV} range. The CMS result \cite{Sirunyan:2018ucr} is shown as the green band, which reflects the combined statistical and systematical uncertainty. The central values of various theoretical predictions (NNLO from \cite{Czakon:2016dgf}, NNLO+EW from \cite{Czakon:2017wor}, and NNLO+NNLL$'$ from \cite{Czakon:2018nun}) are shown in comparison. One can see that there exists a clear gap between the experimental and theoretical results.
It can also be seen that the theoretical predictions in this region strongly depend on $m_t$.
As a result, this discrepancy has a strong impact on the extraction of $m_t$ from kinematic distributions \cite{Sirunyan:2019zvx}, which favors a value of $m_t$ significantly lower than the current world average.
Therefore, it is interesting to ask whether the top quark is indeed lighter than we expect, or there are missing contributions in this region which are not incorporated in the most up-to-date theoretical predictions.

In this Letter, we study a class of higher order QCD corrections in the threshold region $M_{t\bar{t}} \sim 2m_t$. They originate from exchanges of Coulomb-type gluons as well as soft gluons between the top and anti-top quarks. The leading contributions take the form $(\alpha_s/\beta)^n$, where $\beta \equiv \sqrt{1-4m_t^2/M_{t\bar{t}}^2}$ is the velocity of the top and anti-top quarks in the $t\bar{t}$ rest frame. The region $M_{t\bar{t}} \sim 2m_t$ corresponds to $\beta \sim 0$, where the top and anti-top quarks are slowly moving with respect to each other. In the small $\beta$ region, the $1/\beta^n$ contributions (as well as $\ln^n\beta$ ones) are enhanced and lead to large corrections beyond the NNLO(+NNLL$'$) results. These corrections are divergent as $\beta \to 0$, signaling a breakdown of the perturbative expansion in this limit. At NNLO (order $\alpha_s^2$), this does not pose a severe problem since the integration over $M_{t\bar{t}}$ is still convergent, which gives a finite prediction for the averaged differential cross section as shown in Figure~\ref{fig:old_results}. However, at order $\alpha_s^4$ and beyond, even the integration over $M_{t\bar{t}}$ becomes divergent, and one has to perform an all-order resummation to restore the validity of theoretical predictions.
Furthermore, at fixed orders in perturbation theory, $M_{t\bar{t}} \geq 2m_t$ due to phase-space constraint. This is not physical since one knows that bound-state effects can make the mass of the $Q\bar{Q}$ system smaller than $2m_Q$, where $Q$ is a heavy quark. For charm quarks and bottom quarks, such bound-state effects are genuinely non-perturbative. For top-quark pairs, the bound-state effects can be (to a good approximation) studied in perturbation theory, as long as we integrate in a large enough range around the threshold $M_{t\bar{t}} = 2m_t$.

In the following part of this Letter, we present a factorization formula for the partonic differential cross sections valid in the region $M_{t\bar{t}} \sim 2m_t$, which schematically takes the form $d\hat{\sigma} \sim H \times J$. The hard function $H$ describes exchanges and emissions of hard gluons with typical momenta of $\mathcal{O}(M_{t\bar{t}})$, while the potential function $J$ resums exchanges of Coulomb and soft gluons between the top and anti-top quarks.
Note that we are not taking the limit $\sqrt{\hat{s}} \to M_{t\bar{t}}$ here, where $\sqrt{\hat{s}}$ is the partonic center-of-mass energy. As a result, extra hard radiations are allowed and are incorporated in our framework into the hard function $H$. Therefore, our factorization formula is intrinsically different from those in \cite{Hagiwara:2008df, Beneke:2010da, Beneke:2011mq}, where the $\sqrt{\hat{s}} \to M_{t\bar{t}}$ approximation has been employed. Our formalism is similar to that of \cite{Petrelli:1997ge, Kiyo:2008bv}, but takes into account the kinematic dependence of the renormalization and factorization scales, and differs in the treatment of subleading corrections in $\beta$.
Based on our factorization formula, we then present numeric results that are relevant to LHC phenomenology. We show that the resummation effects indeed enhance the $M_{t\bar{t}}$ distribution significantly in the threshold region, and make the theoretical prediction more compatible with the experimental data.

\section{Factorization near threshold}

To set up the stage, we begin with the $t\bar{t}$ invariant mass distribution written as a convolution of the partonic luminosity functions and partonic differential cross sections
\begin{equation}
\frac{d\sigma}{dM_{t\bar{t}}} = \sum_{i,j} \int^1_{\tau} \frac{dz}{z} \frac{\tau}{z} \int d\Theta \, \frac{d\hat{\sigma}_{ij}(z,\mu_f)}{dM_{t\bar{t}} \, d\Theta} \, \ff_{ij}\Big(\frac{\tau}{z},\mu_f\Big) \, ,
\label{eq:fac1}
\end{equation}
where $i,j \in \{q,\bar{q},g\}$ denote partons within the colliding protons; $z \equiv M_{t\bar{t}}^2/\hat{s}$, $\tau \equiv M_{t\bar{t}}^2/s$ and $\mu_f$ is the factorization scale. The parton luminosity functions $\ff_{ij}(y,\mu_f)$ are defined by
\begin{equation}
\ff_{ij}(y,\mu_f) \equiv \int^1_{y} \frac{d\xi}{\xi} \, f_{i/p}(\xi,\mu_f) \, f_{j/p}(y/\xi,\mu_f) \, ,
\end{equation}
where $f_{i/p}$ is the parton distribution function (PDF) of the parton $i$ in the proton $p$. Note that in writing down Eq.~\eqref{eq:fac1}, we have taken into account that the factorization scale $\mu_f$ may depend on kinematic variables (which we collectively denote as $\Theta$) other than $m_t$ and $M_{t\bar{t}}$. This is necessary since we will combine our result with the NNLO result of \cite{Czakon:2016dgf}, in which the scales are correlated with the variable $H_T \equiv \sqrt{p_{T,t}^2+m_t^2} + \sqrt{p_{T,\bar{t}}^2+m_t^2}$, where $p_{T,t}$ and $p_{T,\bar{t}}$ are the transverse momenta of the top and anti-top quarks, respectively.

We are concerned with QCD corrections to the partonic differential cross sections. At the leading order (LO) in $\alpha_s$, only the $q\bar{q}$ and $gg$ channels give non-vanishing contributions
\begin{align}
\frac{d^2\hat{\sigma}_{q\bar{q}}^{(0)}}{dM_{t\bar{t}} \, d\cos\theta_t} &= \frac{2\pi\beta\alpha_s^2(\mu_r)}{M_{t\bar{t}}^3} \frac{C_FC_A}{9} \, c_{q\bar{q},8}(\cos\theta_t) \, \delta(1-z) \, , \nonumber
\\
\frac{d^2\hat{\sigma}_{gg}^{(0)}}{dM_{t\bar{t}} \, d\cos\theta_t} &= \frac{2\pi\beta\alpha_s^2(\mu_r)}{M_{t\bar{t}}^3} \bigg[ \frac{C_F}{32} \, c_{gg,1}(\cos\theta_t) \nonumber
\\
&+ \frac{(C_A^2-4)C_F}{64} \, c_{gg,8}(\cos\theta_t) \bigg] \, \delta(1-z) \, ,
\label{eq:sigma0_hat}
\end{align}
where $\mu_r$ is the renormalization scale and $\theta_t$ is the scattering angle of the top quark in the $t\bar{t}$ rest frame. The coefficients $c_{ij,\alpha}$, with $\alpha=1,8$ labelling the color configuration of the $t\bar{t}$ system, are given by
\begin{align}
c_{q\bar{q},8}(\cos\theta_t) &= \frac{1}{4} \big[ 2 - \beta^2(1-\cos^2\theta_t) \big] \, , \nn
\\
c_{gg,1}(\cos\theta_t) &= \frac{1}{2(1-\beta^2\cos^2\theta_t)^2} \Big[ 4 - 2(1-\beta^2)^2 \nn
\\
&\hspace{-2em} - 2\beta^2(1-\beta^2\cos^2\theta_t) - (1+\beta^2\cos^2\theta_t)^2 \Big] \, ,
\\
c_{gg,8}(\cos\theta_t) &= 2c_{gg,1}(\cos\theta_t) \, \bigg[ \frac{16}{5} - \frac{9}{10} (3-\beta^2\cos^2\theta_t) \bigg] \, . \nn
\end{align}
Note that Eq.~\eqref{eq:sigma0_hat} is proportional to $\beta$ due to the two-body phase space. This leads to the fact that the LO differential cross section approaches zero at the threshold $M_{t\bar{t}} = 2m_t$. This is no longer true at higher orders in $\alpha_s$, essentially because of the $1/\beta^n$ terms mentioned in the Introduction. More precisely, the NLO differential cross section approaches a positive constant in the $\beta \to 0$ limit, while the NNLO and higher order ones diverge in that limit. This behavior makes the perturbative expansion badly convergent in the threshold region. In the following, we analyze the factorization properties of the differential cross sections and resum the large $1/\beta^n$ and $\ln^n\beta$ corrections to all orders in $\alpha_s$. This is the only way to restore the predictive power of perturbative QCD near threshold.

In the threshold limit $M_{t\bar{t}} \to 2m_t$ or $\beta \to 0$, there exist large hierarchies among the energy scales $M_{t\bar{t}}$, $M_{t\bar{t}}\beta$ and $M_{t\bar{t}}\beta^2$. Using the method of regions, we identify the following momentum modes in the $t\bar{t}$ rest frame \cite{Beneke:1997zp}
\begin{align}
\text{hard: } \; &k^\mu \sim M_{t\bar{t}} \, , & \text{potential: } \; &k^0 \sim M_{t\bar{t}}\beta^2 \, , \; \vec{k} \sim M_{t\bar{t}}\beta \, , \nonumber
\\
\text{soft: } \; &k^\mu \sim M_{t\bar{t}}\beta \, , & \text{ultrasoft: } \; &k^\mu \sim M_{t\bar{t}}\beta^2 \, .
\label{eq:modes}
\end{align}
The top and anti-top quarks move very slowly and are non-relativistic objects whose residue momenta correspond to the potential mode. They can interact with each other through soft, ultrasoft and potential gluons which are described by the effective Lagrangian of potential non-relativistic QCD (pNRQCD) \cite{Brambilla:1999xf, Beneke:1999qg}. The hard mode describes fluctuations with typical momenta of $\mathcal{O}(M_{t\bar{t}})$, resulting in Wilson coefficients of effective operators in pNRQCD. We note that there are no collinear modes in our setup. This is related to the fact that we are not taking the limit $\sqrt{\hat{s}} \to M_{t\bar{t}}$, so that the extra emissions are not constrained. As a result, we do not need to employ the soft-collinear effective theory, as opposed to \cite{Beneke:2010da, Beneke:2011mq}.
Using the method of effective field theory (EFT), we derive a factorization formula for the partonic cross sections in the threshold region. The power expansion of pNRQCD is according to the counting $\alpha_s \sim \beta \sim 1/\ln\beta$. Up to the next-to-leading power (NLP), the factorization formula reads
\begin{multline}
\frac{d\hat{\sigma}^{\text{NLP}}_{ij}}{dM_{t\bar{t}} \, d\Theta} = \frac{16\pi^2\alpha_s^2(\mu_r)}{M_{t\bar{t}}^5} \sqrt{\frac{M_{t\bar{t}}+2m_t}{2M_{t\bar{t}}}} \sum_{\alpha} c_{ij,\alpha}(\cos\theta_t)
\\
\times H_{ij,\alpha}(z,M_{t\bar{t}},Q_T,Y,\mu_r,\mu_f) \, J^{\alpha}(E)  + \mathcal{O}(\beta^3) \, ,
\label{eq:fac2}
\end{multline}
where $H_{ij,\alpha}$ are hard functions describing hard gluon exchanges and emissions, which depend on the transverse momentum $Q_T$ and the rapidity $Y$ of the $t\bar{t}$ pair; $J^\alpha$(E) with $E=M_{t\bar{t}}-2m_t$ are potential functions describing exchanges of potential, soft and ultrasoft gluons between the top and anti-top quarks.
Note that at NLP, ultrasoft gluon exchanges among initial-state partons and the $t\bar{t}$ pair do not contribute. As a result there is no soft function in our factorization formula. Such contribution could be present at higher powers in $\alpha_s$ and $\beta$, which is an interesting subject of study in the future.
The prefactors in Eq.~\eqref{eq:fac2} are chosen such that the LO cross sections in the EFT reproduce the exact ones in Eq.~\eqref{eq:sigma0_hat}. These prefactors take into account subleading corrections in $\beta$ at LO, and have significant impacts at higher orders in $\alpha_s$. This, together with the explicit kinematic dependence of the hard function, makes our factorization formula substantially different from those in \cite{Petrelli:1997ge, Kiyo:2008bv}.

The hard functions can be calculated order-by-order in $\alpha_s$ in terms of differential cross sections in the limit $p_{t}=p_{\bar{t}}=P_{t\bar{t}}/2$. For resummation at NLP, we need the hard functions up to the NLO. The LO hard functions are given by
\begin{align}
H^{(0)}_{q\bar{q},1} &= 0 \, , \quad
H^{(0)}_{q\bar{q},8} = \frac{C_A C_F}{9} \, \delta(1-z) \, \delta(Q_T) \, \delta(Y) \, , \nonumber
\\
H^{(0)}_{gg,1} &= \frac{C_F}{32} \, \delta(1-z) \, \delta(Q_T) \, \delta(Y) \, , \nonumber
\\
H^{(0)}_{gg,8} &= \frac{(C_A^2-4)C_F}{64} \, \delta(1-z) \, \delta(Q_T) \, \delta(Y) \, .
\end{align}
We have calculated the NLO corrections to the hard functions analytically, which were not known in the literature. They receive contributions from both virtual exchanges and real emissions of gluons. The virtual and real contributions are separately infrared divergent. The divergences cancel when combining the two together with the PDF counterterms. After the cancellation, we obtain finite NLO hard functions involving singular distributions of $1-z$, $Q_T$ and $Y$. The results are more complicated than those in the literature \cite{Petrelli:1997ge, Kiyo:2008bv} since we need to keep the dependencies on kinematic variables in order to use the $H_T$-based renormalization and factorization scales.
The potential function $J^\alpha(E)$ is related to the imaginary part of the pNRQCD Green function $G^\alpha(\vec{r}_1,\vec{r}_2;E)$ of the $t\bar{t}$ pair at origin. It receives contributions from potential, soft and ultrasoft modes.
The explicit expressions for the Green function up to NLP can be found in, .e.g., \cite{Beneke:2011mq, Ju:2019lwp}.
The finite width of the top quark can be taken into account by the replacement $E \to E+i\Gamma_t$ in the potential function. Such a treatment is valid at NLP, but additional considerations are required at higher powers (we count $\Gamma_t/m_t \sim \beta^2$) \cite{Hoang:1999zc, Beneke:2003xh, Beneke:2004km, Beneke:2010mp, Beneke:2017rdn}.

Combining the hard and potential functions, we can now produce NLP resummed predictions for the $M_{t\bar{t}}$ distribution near threshold. These can be further combined with fixed-order ones via the matching formula
\begin{equation}
\frac{d\sigma^{\text{(N)NLO+NLP}}}{dM_{t\bar{t}}} = \frac{d\sigma^{\text{NLP}}}{dM_{t\bar{t}}} + \frac{d\sigma^{\text{(N)NLO}}}{dM_{t\bar{t}}} - \frac{d\sigma^{\text{(n)nLO}}}{dM_{t\bar{t}}} \, ,
\label{eq:matching}
\end{equation}
where we have used ``nLO'' and ``nnLO'' to denote fixed-order expansions of the NLP resummed result \eqref{eq:fac2} up to the second and third order in $\alpha_s(\mu_r)$. Their differences with respect to the full NLO and NNLO results are higher-power terms in $\beta$ that we want to take into account through the matching procedure.

\section{Numeric results}

We now present the numeric results based on our factorization formula \eqref{eq:fac2} and the matching formula \eqref{eq:matching}. Throughout our calculation we take $m_t=\unit{172.5}{\GeV}$, $\Gamma_t=\unit{1.4}{\GeV}$ and use the NNPDF3.1 NNLO PDFs with $\alpha_s(m_Z)=0.118$. Since the small-$\beta$ resummation is valid only at low $M_{t\bar{t}}$, we restrict ourselves to $M_{t\bar{t}} \in \unit{$[300,380]$}{\GeV}$ in this work, which is the region most sensitive to $m_t$. Following \cite{Czakon:2017wor, Czakon:2018nun}, we choose the default values of the renormalization scale $\mu_r$ and the factorization scale $\mu_f$ to be $H_T/4$, the corresponding results are regarded as the central values. We then vary the scales simultaneously up and down by a factor of 2 to estimate the remaining theoretical uncertainties. Note that with such a procedure, the error bands are not necessarily symmetric around the central values.

\begin{figure}[t!]
\centering
\includegraphics[width=0.9\linewidth]{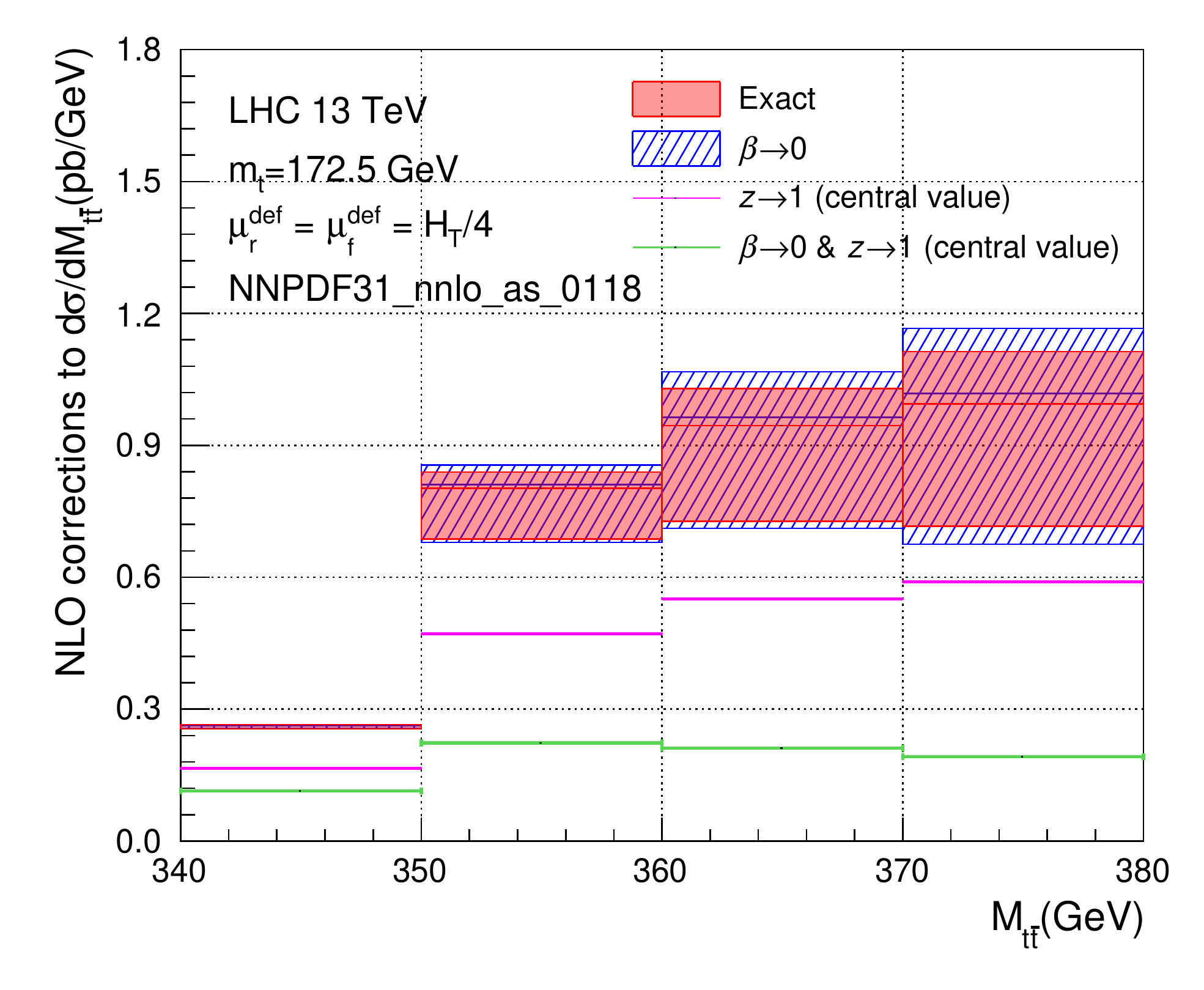}
\vspace{-4ex}
\caption{\label{fig:nlo_vs_approx}The exact NLO correction and its various approximations in the range \unit{[340-380]}{\GeV}.}
\end{figure}

First of all, in Fig.~\ref{fig:nlo_vs_approx} we show the exact NLO correction computed using \texttt{MCFM}~\cite{Campbell:2010ff}, in comparison with its approximations in various kinematic limits. Such a comparison is important to assess the validity of the approximations within the phase-space region of interest. Fig.~\ref{fig:nlo_vs_approx} has a few significant implications we'd like to elaborate.
\textbf{1)} The $\beta \to 0$ result corresponds to the second term in the fixed-order expansion of our NLP formula \eqref{eq:fac2}. One can see that it provides an excellent description of the full NLO correction in the range \unit{$[340,380]$}{\GeV}. This demonstrates that our small-$\beta$ approximation is indeed valid, and supports the resummation performed in this work.
\textbf{2)} The $z \to 1$ result corresponds to the ``soft'' or ``threshold'' limit $\sqrt{\hat{s}} \to M_{t\bar{t}}$ considered in, e.g., \cite{Ahrens:2010zv, Pecjak:2016nee}. It should be stressed that the concept of ``soft'' or ``threshold'' there has a completely different meaning than those used in this work. We observe that, while not perfect, the $z \to 1$ limit provides a reasonable approximation to the full result in the low $M_{t\bar{t}}$ region. It is known that this limit works better in the high $M_{t\bar{t}}$ region \cite{Ahrens:2010zv, Pecjak:2016nee}.
\textbf{3)} Finally, the double limit $\beta \to 0$ \emph{and} $z \to 1$ corresponds to the ``soft'' limit considered in \cite{Hagiwara:2008df, Kiyo:2008bv}. One can see that this approximation does \emph{not} capture the dominant contribution at NLO. This is essentially the reason why we do not consider such a ``soft'' resummation in this work.

\begin{figure}[t!]
\centering
\includegraphics[width=0.9\linewidth]{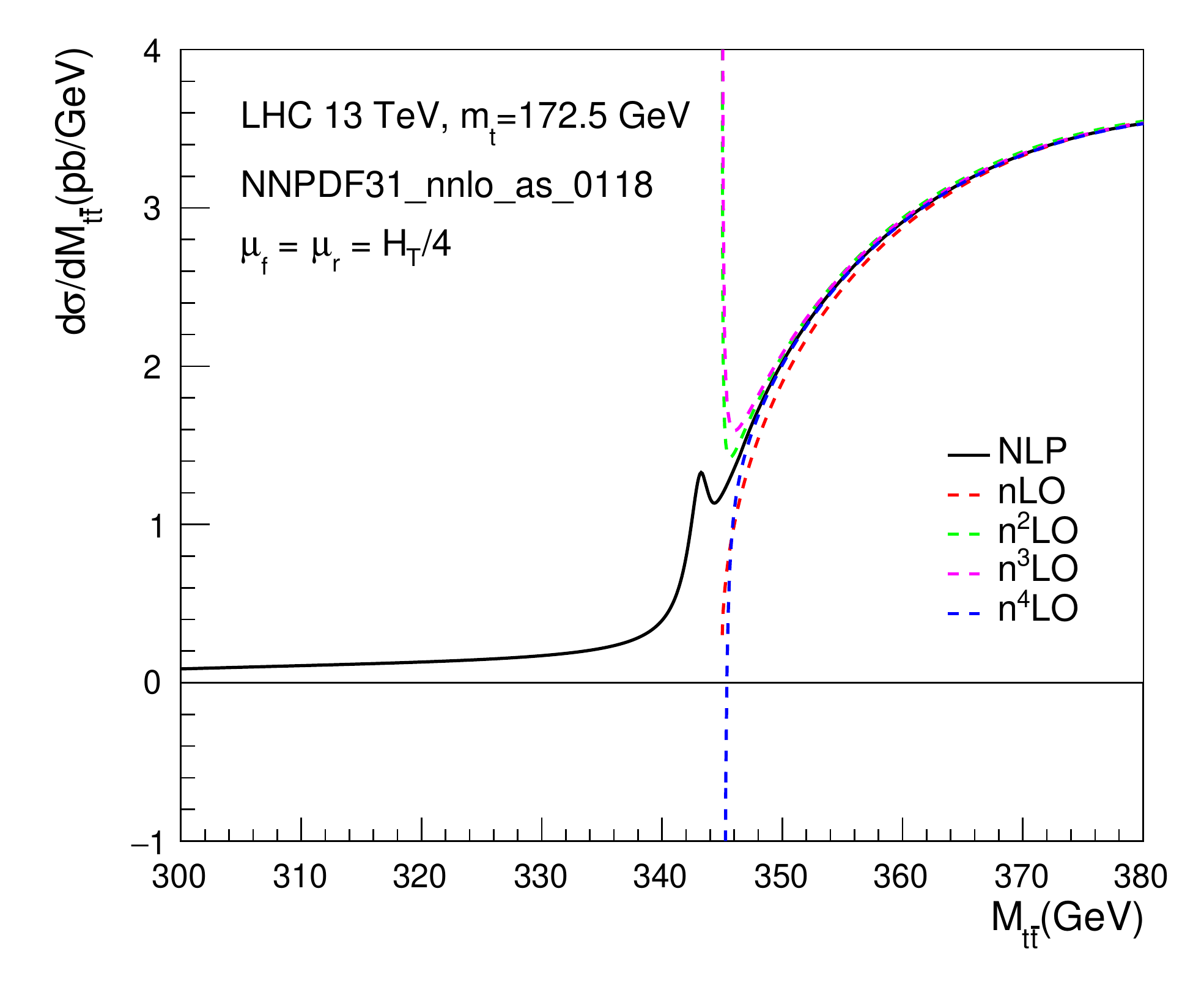}
\vspace{-4ex}
\caption{\label{fig:nlp_expansion}The NLP resummed result and its fixed-order expansion.}
\end{figure}

We now turn to compare the NLP resummed result Eq.~\eqref{eq:fac2} and its fixed-order expansion. As in Eq.~\eqref{eq:matching}, we label the expansion up to the $i$-th term ($i=0,1,2,\ldots$) as n$^{i}$LO. Here the lower-case `n' implies that these are approximations to the full N$^{i}$LO results in the $\beta \to 0$ limit. Note that the n$^0$LO result equals exactly to the full LO result due to the prefactors in Eq.~\eqref{eq:fac2}. We show such a comparison in Fig.~\ref{fig:nlp_expansion}, from which we can draw several important conclusions.
\textbf{1)} The fixed-order expansion converges rather quickly when $M_{t\bar{t}}$ is much larger than $2m_t$. However, when $M_{t\bar{t}}$ approaches the threshold, the behavior becomes out-of-control. In particular, the n$^3$LO result tends to $+\infty$ while the n$^4$LO one tends to $-\infty$ in the $\beta \to 0$ limit ($M_{t\bar{t}} \to \unit{345}{\GeV}$).
\textbf{2)} The singularity at $\beta = 0$ is regularized by the resummation effects, and we obtain a finite prediction near $M_{t\bar{t}} = 2m_t$ with the NLP resummation formula. One also finds that in the resummed result the region $M_{t\bar{t}} < 2m_t$ is allowed due to bound-state effects. The shape of the NLP curve for $M_{t\bar{t}} < 2m_t$ depends crucially on the top quark width. However, we have checked that the integrated cross section in the \unit{$[300,380]$}{\GeV} range is insensitive to the value of $\Gamma_t$.
\textbf{3)} As a final implication of Fig.~\ref{fig:nlp_expansion}, we note that the NLP curve almost overlaps with the nLO and n$^2$LO curves for $M_{t\bar{t}} > \unit{360}{\GeV}$. This means that in this region, the matched (N)NLO+NLP results of Eq.~\eqref{eq:matching} are governed by the fixed-order (N)NLO calculations. Effectively, this shows that we are not applying the small-$\beta$ resummation to regions where one may worry about the break down of the EFT description. As a matter of fact, we have checked that the dominant beyond-NNLO correction comes from the region $M_{t\bar{t}} < \unit{350}{\GeV}$, where $\beta < 0.17$ and pNRQCD is perfectly applicable.

\begin{figure}[t!]
\centering
\includegraphics[width=0.9\linewidth]{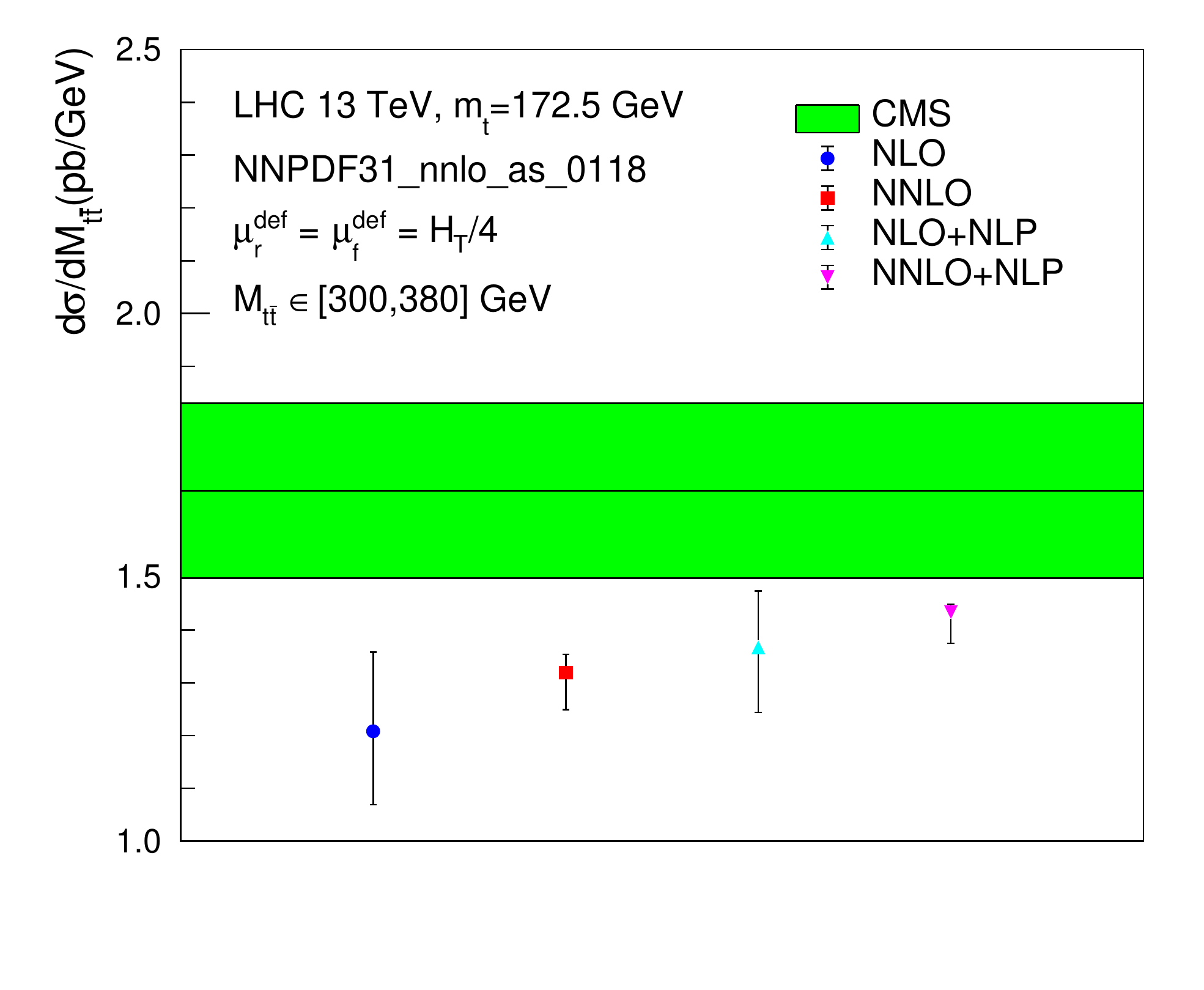}
\vspace{-7ex}
\caption{\label{fig:our_results}The averaged $t\bar{t}$ invariant mass distribution in the \unit{[300,380]}{\GeV} range. The CMS result \cite{Sirunyan:2018ucr} is shown as the green band. The various theoretical predictions are shown in comparison, with NNLO+NLP being our best prediction.}
\end{figure}

\begin{table}[t!]
\centering
\begin{tabular}{|c|c|}
\hline
CMS & $1.664^{+0.166}_{-0.166}$
\\ \hline
NLO & $1.208^{+0.150}_{-0.139}$
\\ \hline
NNLO & $1.319^{+0.035}_{-0.070}$
\\ \hline
NLO+NLP & $1.367^{+0.107}_{-0.124}$
\\ \hline
NNLO+NLP & $1.434^{+0.014}_{-0.060}$
\\ \hline
\end{tabular}
\caption{\label{tab:our_results}The differential cross sections (in unit of pb/GeV) entering Fig.~\ref{fig:our_results}.}
\end{table}

Finally, we match our resummed calculation to the NLO and NNLO results according to Eq.~\eqref{eq:matching}. The NLO results are computed using \texttt{MCFM} and the NNLO results are obtained from \cite{Czakon:2016dgf, Catani:2019hip, Britzger:2012bs, Czakon:2017dip}. The NLO+NLP and NNLO+NLP results are shown in Fig.~\ref{fig:our_results} and Tab.~\ref{tab:our_results} together with the NLO and NNLO ones, compared against the CMS measurement. We find that the resummation effects enhance the NNLO differential cross section by about 9\%, and make the theoretical prediction more compatible with experimental data. As discussed in the Introduction, this enhancement will have significant impacts on the extraction of the top-quark mass from kinematic distributions. Although we can't repeat the analysis of \cite{Sirunyan:2019zvx}, we can estimate the impacts by studying the $m_t$-dependence of the NLO
(used in the fit of \cite{Sirunyan:2019zvx}) and NNLO+NLP predictions for the \unit{[300,380]}{\GeV} range. For example, the central value of the NNLO+NLP differential cross section for $m_t = \unit{172.5}{\GeV}$ is given by \unit{1.434}{\picobarn\per\GeV}. To achieve the same differential cross section using the NLO calculation, one needs to lower the value of $m_t$ down to about \unit{171}{\GeV}.
This roughly explains the outcome of \cite{Sirunyan:2019zvx}, and shows that using our prediction in the fit will lead to a shift of about \unit{1.5}{\GeV} resulting in a value closer to the world average.

\section{Conclusion}

In this Letter, we studied the $t\bar{t}$ invariant-mass distribution near the $2m_t$ threshold. In this region, there exist a small gap between the most up-to-date theoretical predictions and the experimental measurements by the CMS collaboration at the LHC. This leads to a tension between the values of $m_t$ determined from direct measurements and from fitting kinematic distributions in $t\bar{t}$ production.
We show that higher-order non-relativistic effects lead to large corrections to the differential cross section in the threshold region, which were not included in the state-of-the-art theoretical predictions. We derive a factorization formula to resum such corrections to all orders in the strong coupling, and calculate necessary ingredients to perform the resummation at next-to-leading power. We combine the resummation with NLO and NNLO results, and present numeric results relevant for LHC phenomenology. We find that the resummation effect increases the differential cross section in the range $M_{t\bar{t}} \in \unit{[$300,380$]}{\GeV}$ by about 9\%. This makes the theoretical prediction more compatible with experimental data, and leads to a shift of about \unit{1.5}{\GeV} on the extracted top-quark mass from kinematic distributions, resulting in a value better consistent with the world average.

Our results can be easily combined with soft gluon resummation of \cite{Ahrens:2010zv, Pecjak:2016nee, Pecjak:2018lif, Czakon:2018nun} and with electroweak corrections of \cite{Bernreuther:2010ny, Pagani:2016caq, Czakon:2017wor, Czakon:2019txp}. This allows a precision prediction for the $t\bar{t}$ invariant-mass spectrum across the whole phase space. Our framework can be extended to study the $t\bar{t}+\text{jets}$ production process, which is also being used to extract the top quark mass \cite{Alioli:2013mxa, Aad:2019mkw}.

\begin{acknowledgments}
L.~L.~Yang would like to thank A.~Mitov for useful discussions. This work was supported in part by the National Natural Science Foundation of China under Grant No. 11975030, 11635001 and 11575004. W.-L.~Ju was supported in part by the China Postdoctoral Science Foundation under Grant No. 2017M610685.
\end{acknowledgments}

\end{document}